\documentclass[prl,twocolumn,superscriptaddress,english,footinbib]{revtex4-2}
\usepackage{amsmath,amssymb,mathrsfs}
\usepackage{amsfonts,bm}
\usepackage{amsmath}
\usepackage{amssymb}
\usepackage{amsthm}
\usepackage{graphicx}
\usepackage{graphics}
\usepackage{subfigure}
\usepackage{color}
\usepackage{bm}
\usepackage{hyperref}
\usepackage{rotating}
\usepackage{comment}
\usepackage[colorinlistoftodos]{todonotes}
\usepackage{esint}
\usepackage{xfrac}
\usepackage{natbib}
\usepackage{upgreek}

\newcommand{\be}{\begin{equation} }
\newcommand{\ee}{\end{equation} }
\newcommand{\ba}{\begin{eqnarray} }
\newcommand{\ea}{\end{eqnarray} }

\newcommand{\bpm}{\begin{pmatrix}}
\newcommand{\epm}{\end{pmatrix}}
\newcommand{\bmm}{\begin{matrix}}
\newcommand{\emm}{\end{matrix}}

\newcommand{\la}{\label}
\newcommand{\p}{\partial}

\newcommand{\bea}{\begin{eqnarray}}
\newcommand{\eea}{\end{eqnarray}}

\makeatother

\usepackage{babel}

\begin{document}


\title{Kardar-Parisi-Zhang Universality at the Edge of Laughlin States}

\author{Gustavo M. Monteiro}
\affiliation{Department of Physics and Astronomy, College of Staten Island, CUNY, Staten Island, NY 10314, USA}
\author{Dylan Reynolds}
\affiliation{Department of Physics, City College, City University of New York, New York, NY 10031, USA }
\author{Paolo Glorioso}
\affiliation{Department of Physics, Stanford University, Stanford CA 94305, USA}
\author{Sriram Ganeshan}
\affiliation{Department of Physics, City College, City University of New York, New York, NY 10031, USA }
\affiliation{CUNY Graduate Center, New York, NY 10031}

\date{\today}


\begin{abstract}

In this letter, we investigate the dissipative dynamics at the edge of  Laughlin fractional quantum Hall (FQH) states starting from the hydrodynamic framework of the composite Boson theory recently developed in arXiv:2203.06516. Critical to this description is the choice of boundary conditions, which ultimately stems from the choice of hydrodynamic variables in terms of condensate degrees of freedom. Given the gapped nature of bulk, one would expect dissipation effects to play an important role only near the FQH edge. Thus, one envisions a scenario where the bulk hydro equations remain unmodified, while  the dissipation effects are introduced at the edge via boundary conditions. We have recently shown that the anomaly requirements fix the boundary conditions of the FQH fluid to be no-penetration and no-stress boundary conditions. 
In this work, we introduce energy dissipation in the no-stress boundary condition leading to charge diffusion at the boundary. The resulting dissipative edge dynamics is quite rigid from a hydro perspective, as it has to preserve the edge charge continuity and the anomaly structure. We show that the diffusive edge dynamics with fluctuation-dissipation relations within a power counting scheme belong to the Kardar-Parisi-Zhang universality class.

\end{abstract}


\maketitle


{\bf Introduction:} The superfluid framework of Laughlin fractional quantum Hall (FQH) states was originally put forward by Girvin and Macdonald \cite{girvin1987off}. This framework draws an analogy between the Laughlin wave function and a Bose condensate, interpreting the FQH state as a charged superfluid of composite bosons similar to superfluid helium and superconductivity~\cite{helium-book, Abrikosov1957TheMP, stone1990superfluid}. Zhang, Hansson, Kivelson~\cite{zhang1989effectivetheory,zhang1992csgl} and Read~\cite{read1989orderparameter} further developed this composite boson picture of Laughlin states in the form of the Chern-Simons-Ginzburg-Landau (CSGL) theory. The CSGL theory consists of a charged condensate coupled to a statistical Chern-Simons gauge field. The dynamics of this charged composite boson condensate can be expressed in terms of hydrodynamic variables, with governing equations taking the form of charge continuity and momentum conservation subject to an additional constitutive relation known as the Hall constraint. This Hall constraint relates the charge density and magnetic field to the fluid vorticity~\cite{stone1990superfluid, abanov2013fqhe}. 

Our recent work~\cite{monteiro2022topological, monteiro2023coastal} has demonstrated that the anomaly requirements determine the boundary conditions of the FQH fluid at a hard wall. Specifically, we found that the boundary conditions must be no-penetration and no-stress conditions, which is a crucial distinction when we consider the higher gradient quantum pressure (Madelung) terms. In two-dimensional condensates, the quantum pressure terms, which involve two derivatives of the density, can be expressed as an odd viscosity term with a specific choice of the superfluid velocity~\cite{abanov2013fqhe, geracie2014effective, monteiro2022topological, monteiro2021hamiltonian}. With this velocity field, the hydrodynamic equations become first-order in velocity gradients. Furthermore, our work in Ref.~\cite{monteiro2022topological} has revealed that the no-stress boundary condition is a dynamical equation at the edge, which can be derived from a chiral boson action non-linearly coupled to the condensate density at the edge \footnote{In the pure Chern-Simons (CS) framework of the FQH edge the bulk has no dynamics, and the CS term in the bulk only fixes the edge algebra through gauge invariance. The edge dynamics are subsequently added by hand~\cite{wen1990compressibility}. In contrast,  within the first-order hydro framework, the bulk and boundary dynamics are  intricately tied together by the no-stress boundary conditions, manifesting boundary layer structure and naturally resulting in a dynamical chiral boson theory that is coupled to the matter density at the edge.}.

These hydrodynamic equations for the FQH state, when linearized, give rise to two chiral modes at the edge: a non-dispersive Kelvin mode and a dispersing chiral boson mode, both propagating in the same direction. The coexistence of these two chiral modes in the Laughlin state may initially appear paradoxical. However, in our work ~\cite{monteiro2023coastal}, we demonstrated that the Kelvin mode does not contribute to edge charge transport. On the other hand, the dispersing chiral boson mode does exhibit consistent gauge anomaly-induced chiral edge dynamics.

The FQH system we have discussed so far corresponds to a non-dissipative superfluid that can be described variationally in both bulk and boundary regions. However, in realistic experimental setups, the FQH edge is expected to dissipate \cite{kane1994randomness,kane1995impurity}. At low temperatures, the bulk remains non-dissipative, but the gapless edge modes may potentially relax. Additionally, as a consequence of dissipation and assuming proximity to thermal equilibrium, the edge is subject to thermal noise. Recent work by one of the co-authors \cite{delacretaz2020breakdown} has shown that the interplay between thermal fluctuations and the chiral anomaly leads to \emph{superdiffusive} damping of the edge mode, as opposed to diffusive damping predicted by mean-field considerations. Furthermore, this superdiffusive scaling is a manifestation of the celebrated Kardar-Parisi-Zhang (KPZ) universality class \cite{kardar1986dynamic}. KPZ universality has been appreciated to emerge in a disparate variety of contexts in hydrodynamics, from one-dimensional systems with momentum conservation \cite{forster1977large}, to integrable \cite{vznidarivc2011spin,wei2022quantum} and non-integrable \cite{das2020nonlinear} spin chains.

In this letter, we employ the hydrodynamic framework to incorporate dissipation at the edge of the FQH state. To achieve this, we relax the conservation of energy to account for dissipation occurring at the edge of the FQH fluid. In the energy-conserving case, the no-stress boundary condition can be recast as a continuity equation  for the edge density given by,
\begin{align}
    \left[\p_t(\sqrt{ n})+\p_x(\sqrt{ n}\, v_x)\right]\Big|_{y=0}&=0 \,.
\end{align}
The emergence of this $U(1)$ symmetry for the edge density ($\sqrt{n}|_{y=0}$) is a natural outcome of the boundary layer physics, as discussed in~\cite{monteiro2022topological}. Moreover, the bulk dynamics determines the boundary constitutive relation $v_x[n,\partial_x n, ..]$, which in turn gives rise to effective edge dynamics solely described by the fluctuations of the condensate density at the boundary~\cite{monteiro2023coastal}. Within this setup, the edge relaxation can be introduced through a dissipative term added to the edge current  $J(x,t)=\left(\sqrt{n}v_x|_{y=0}+J_{\text{diss}}\right)$, where the form of $J_{\text{diss}}$ is constrained by the second law of thermodynamics. This edge dissipation can also be viewed as modifying the no-stress boundary condition. The presence of such a constrained dissipation naturally leads to the diffusive dynamics of the ballistic chiral front within linear order. When nonlinearities and fluctuations are introduced, local hydrodynamics breaks down and the dynamics of the front become superdiffusive as discussed above. In contrast to Ref.  \cite{delacretaz2020breakdown}, 
where the 1D anomalous hydrodynamics with only one conserved quantity was the starting point, here the constitutive relations arise from the full 2D system, which fixes all the edge transport coefficients.

 This paper is organized as follows. We begin with the fluid description of the Laughlin states starting from the CSGL action. We then introduce the boundary conditions at a hard wall consistent with the gauge anomaly in the dissipationless case. We then modify the boundary conditions to include dissipation at the edge, derive the equation of edge charge diffusion, and examine the linearized system to obtain the surface dispersion. Finally, we discuss the emergence of KPZ universality at the edge of the Laughlin state upon including the non-linearities and noise terms.

{\bf Hydrodynamic Description of the Laughlin FQH state:} 
The CSGL theory consists of a charged bosonic matter field $\Phi$ coupled to an external electromagnetic field $A_\mu$, along with a statistical Chern-Simons gauge field $a_\mu$. Here, $\mu$ denotes both temporal and spatial components of the vector fields, that is, $\mu=0,1,2$. The variational principle for this model can be expressed as follows:
\begin{align}
S_{\text{CSGL}}&=\int d^3x\left[i\hbar\Phi^\dagger D_t\Phi - \frac{\hbar^2}{2m}|D_i\Phi|^2-V(|\Phi|^2)\right.\nonumber
\\
&\left.-\frac{\hbar\nu}{4\pi}\epsilon^{\mu\lambda\kappa} a_\mu\p_\lambda a_\kappa\right], \la{S-CSGL}
\end{align}
where $\nu$ represents the filling factor, and $D_\mu=\partial_\mu+i\left(\frac{e}{\hbar}A_\mu+a_\mu\right)$ denotes the covariant derivative, with $e$ and $m$ representing the charge and effective mass of the  quasiparticle, respectively. The equations of motion derived from this action can be recast as hydro equations using the Madelung transformation $\Phi=\sqrt{n}\,e^{i\theta}$ with fluid velocity defined as $v_i=\frac{\hbar}{m}\left(\p_i\theta+a_i+\frac{e}{\hbar} A_i+\frac{\epsilon_{ij}}{2n}\p^jn\right)$. This choice of velocity field leads to the following first-order hydro equations (see appendix for details of this calculation),
\begin{align}
\partial_t n + \partial_i\left(n v_i\right) &= 0, \label{eq:chargecon}\\
\partial_t v_i + v_j\partial_j v_i +\frac{eB}{m}\epsilon_{ij}v_j - \frac{1}{m n}\partial_j T_{ij}&=0,\label{eq:momcon}\\
n-\frac{\nu eB}{2\pi\hbar}+\frac{\nu}{4\pi}\partial_i\left(\frac{\partial_i n}{n}\right) + \frac{\nu m}{2\pi\hbar}\epsilon_{ij}\partial_i v_j &=0. \label{eq:hallcon}
\end{align}
Equation~(\ref{eq:chargecon}) is the continuity equation, Eq.~(\ref{eq:hallcon}) is the Hall constraint that pins the superfluid vorticity to fluctuations of the condensate density, and Eq.~(\ref{eq:momcon}) represents the modified Euler equation, with the stress tensor given by
\begin{align}
T_{ij} = \left(V-nV'-\frac{\pi\hbar^2n^2 }{\nu m}\right)\delta_{ij} - \frac{\hbar n}{2}\left(\epsilon_{ik}\partial_k v_j + \epsilon_{jk}\partial_i v_k\right).\label{eq:tij}
\end{align}
Here the potential term is taken to be a simple short-range contact repulsion with strength $g$, that is, $V(n)=g(n-\frac{\nu eB}{2\pi\hbar})^2$, which pins the condensate density to $\nu eB/(2\pi\hbar)$. The last term is of the form of an odd viscosity tensor with $-\hbar n/2$ as the $\nu$ independent hydrodynamic odd viscosity, for a detailed derivation and discussion of the above see~\cite{monteiro2022topological}. 

{\bf Anomaly compatible hydro boundary conditions:}  In this work, we take the fluid domain to be $y \leq 0$, to model an FQH fluid in the lower half-plane.  Since the stress tensor~(\ref{eq:tij}) is first order in velocity gradients, we need to impose two boundary conditions. Before considering any dissipation, the boundary conditions can be obtained by enforcing charge and energy conservation. From the charge conservation requirement, we get no-penetration condition $v_y\big|_{y=0}=0$. The second boundary condition can be obtained following the energy conservation requirement,
\begin{align}
\frac{dE}{dt} = \frac{d}{dt}\int d^2 x\, \mathcal{H}=0,
\end{align}
where the energy density $\mathcal{H}$ is given by
\begin{align}
\mathcal{H} = \frac{m n}{2} v_i v_i + V(n) .
\end{align}
From equations (\ref{eq:chargecon}), (\ref{eq:momcon}), and (\ref{eq:tij}) we obtain
\begin{align}
\frac{dE}{dt} &=-\int d^2 x\,\partial_i\left(\mathcal{H}v_i + T_{ij}v_j \right)=-\int dx \left(T_{xy}v_x\right)\Big|_{y=0}. \label{eq:edgeenergy}
\end{align}
In the above equation, we have used no-penetration boundary condition $v_y\big|_{y=0}=0$ and that the fields vanish at infinity \footnote{Since the stress tensor $T_{ij}$ is symmetric, we have used $T_{xy}=T_{yx}$}. From Eq.~(\ref{eq:edgeenergy}) it is clear that if energy is to be conserved, we must either have the no-slip boundary condition or the no-stress boundary condition. As shown in Ref.~\cite{monteiro2022topological}, only the no-stress condition is compatible with the gauge anomaly. In fact, the no-stress condition can be written in the form of a continuity equation which can be interpreted as an emergent U(1) symmetry if one identifies the edge density to be $\sqrt{n}|_{y=0}$. Rewriting the no-stress condition using the charge continuity equation and no-penetration boundary condition, we have
\begin{align}
T_{xy}\Big|_{y=0} = \hbar \sqrt{n}\Big[ \partial_t \left(\sqrt{n}\right)+\partial_x\left(\sqrt{n}v_x\right)\Big]_{y=0}=0 . \label{eq:tyx}
\end{align}
We see that the no-stress boundary condition  implies that edge charge density $\sqrt{n}|_{y=0}$ is conserved.

{\bf Edge Dissipation: } We now adapt the previous analysis to introduce dissipation at the edge by modifying the no-stress boundary condition~\footnote{In fluid dynamics, the no-penetration condition is typically unchanged regardless of the dissipative nature of the fluid. Thus, we do not change the no-penetration  condition in order to introduce dissipation.} while retaining the structure of the edge continuity equation. Since the bulk is gapped, it is reasonable to assume that the dissipation is confined near the boundary and the bulk equations are unchanged. For dissipation to occur we must satisfy $\frac{dE}{dt} \leq 0$, and the natural way to enforce this condition is to have the integrand of Eq. (\ref{eq:edgeenergy}) be a quadratic form. One possible choice of boundary condition is the partial slip boundary condition~\cite{levitov2016electron}, where $T_{xy}\propto v_x$. However, it is straightforward to check that the partial slip boundary condition explicitly breaks the structure of the edge continuity equation. In fact, no function of the boundary fields $n$ and $v_x$ alone will work.  

The choice of boundary condition which preserves the edge continuity is of the form,
\begin{align}
T_{xy}\Big|_{y=0} = -\hbar\sqrt{n}\, \partial_x J_{\text{diss}}\Big|_{y=0}, \label{eq:txytemp}
\end{align}
where $J_{\text{diss}}$ is some general local function of $n$, $v_x$ and their gradients in the $x$ direction. Therefore, the dynamical form of the modified no-stress boundary condition can be written as,
\begin{align}
\Big[ \partial_t \left(\sqrt{n}\right)+\partial_x\left(\sqrt{n}v_x +J_{\text{diss}}\right)\Big]_{y=0} = 0. \label{eq:tempcont}
\end{align}
Upon inserting the form (\ref{eq:txytemp}) into (\ref{eq:edgeenergy}) and integrating by parts, we obtain
\begin{align}
\frac{dE}{dt}= -\hbar\int dx\, J_{\text{diss}} \,\partial_x \left(\sqrt{n}v_x\right)\Bigg|_{y=0}\leq0.
\end{align}
The dissipation occurs when the integrand is a quadratic form. Thus, the most general form for $J_{\text{diss}}$ is given by
\begin{align}
J_{\text{diss}}= \sum_{r=0}^{\infty}(-1)^{r} \,b_r\,\partial_x^{2r+1} \left(\sqrt{n}v_x\right)\Big|_{y=0}, \label{eq:qdef}
\end{align}
with $b_r\geq 0$, for all $r$.  For the rest of this work, we only consider the minimal case with $b_0=\Gamma^2$ and $b_r=0$ for all $r>1$ as it is the most dominant term.

{\bf Linear Analysis:} In this section we investigate how dissipation at the edge modifies the linear mode structure of the system, and how the modified boundary condition (\ref{eq:txytemp}) leads to charge diffusion near the edge. For a detailed analysis of this linearized FQH fluid system without any dissipation, see Ref.~\cite{monteiro2023coastal}. Here we only outline the key steps and highlight the important differences that arise due to dissipation. We linearize around a constant background density $n=\frac{\nu e B}{2\pi\hbar}(1+\rho)$, where $\rho$ represents the small density perturbation. The velocity fields have no background flow and can be expressed as $v_x=u$ and $v_y=v$. Using the interaction-dependent sound velocity $c^2=\ell_B^2\omega_B^2+\frac{\omega_B \nu g}{2\pi\hbar}$, the governing equations (\ref{eq:chargecon})-(\ref{eq:tij}) can be written as follows,
\begin{align}
 &\partial_t\rho + \partial_x u+\partial_y v =0 ,\\
&\partial_t u + c^2\partial_x\rho + \omega_B^2\left(1+\frac{\ell_B^2}{2}(\p_x^2+\p_y^2)\right)v =0 ,\\
&\partial_t v + c^2\partial_y\rho - \omega_B^2\left(1+\frac{\ell_B^2}{2}(\p_x^2+\p_y^2)\right)u =0 ,\\
&(\p_x v-\p_y u)+\omega_B^2\left(1+\frac{\ell_B^2}{2}(\p_x^2+\p_y^2)\right)\rho=0.
\end{align}
Here we have  defined the cyclotron frequency and magnetic length scale as $\omega_B=eB/m$ and $\ell_B = \sqrt{\hbar/(eB)}$, respectively. The linearized no-penetration and modified no-stress boundary conditions at $y=0$ become,
\begin{align}
v=0,\quad 
\partial_t\rho + 2\partial_x u-2\Gamma^2\partial_x^2 u=0.\label{eq:linstress}
\end{align}
The bulk flow admits plane wave solutions of the form $(\rho, u, v) \propto e^{i\bm{q}\cdot\bm{x}-i\omega t}$ with dispersion
\begin{align}
	\omega=\pm \sqrt{c^2q^2+\omega_B^2\left(1-\tfrac{1}{2}q^2\ell_B^2\right)^2}\,,
 \label{eq:disp-bulk}
\end{align}
The edge flow is described by modes localized near the rigid interface indicating the following ansatz $e^{s y +i kx -i\omega t}$.  By making the replacement $(q_x, q_y)\to (k, -is)$, the dispersion relation (\ref{eq:disp-bulk}) can be extended to incorporate the edge, resulting in a quartic polynomial in $s$. The two solutions with positive real parts are denoted $s_1$ and $s_2$ and are related by $s_2=s_1^*$. The edge solutions are then written as a linear superposition of these two modes. Note that to prevent inter-band mixing $s$ must retain an imaginary piece, which further constrains the sound velocity $c^2\leq 2\ell_B^2\omega_B^2$~\cite{monteiro2023coastal}.

Upon applying the no-penetration boundary condition, we immediately see two solutions. The first solution is given by $(\rho,u,v)=(\rho_K, c \rho_K, 0)$, with corresponding dispersion $\omega=ck$. This is known as the coastal Kelvin mode, in reference to the coastal wave solutions in the shallow water model. The Kelvin modes are naturally non-dispersive, $\rho_K(x,y,t)=\rho_K(x-ct,y)$, which when combined with (\ref{eq:linstress}) gives
\begin{align}
\partial_{\xi}\rho_\text{K}-2\Gamma^2\partial_{\xi}^2\rho_\text{K}=0,
\end{align}
where $\xi=x-ct$ is the boosted coordinate. The only bounded solution is a constant, and requiring that the fields must vanish at $x=\pm \infty$, we obtain the density fluctuation to vanish at the edge $\rho_K(\xi,0)=0$. In other words, the Kelvin mode does not contribute to energy dissipation and therefore will not participate in heat transport at the edge.

The second solution corresponds to the chiral boson mode, and the dispersion relation is obtained by imposing the modified no-stress boundary condition, which now includes the dissipative term. By examining the structure of the eigenvectors for this mode, we can establish a relationship between the edge flow along the boundary and the edge density in the long wavelength limit,
\begin{align}\label{uBC}
u_{\text{CB}}(x,0,t)=c \left(1+\frac{\ell_B^2}{4}\partial_x^2\right) \rho_{\text{CB}}(x,0,t) + \cdots.
\end{align}
For detailed calculations see \cite{monteiro2023coastal} and the supplementary material therein. Upon substitution into the linearized no-stress boundary condition (\ref{eq:linstress}) we arrive at an equation that governs the leading order dynamics of the chiral boson edge mode in the presence of dissipation,
\begin{align}
\left(\partial_t + 2c\partial_x - 2c\Gamma^2\partial_x^2 + \frac{c\ell_B^2}{2}\partial_x^3 + \cdots\right)\rho_{\text{CB}}(x,0,t) = 0. \label{eq:diffusion}
\end{align}
From this, we clearly see that the inclusion of dissipation at the boundary induces edge charge diffusion, stemming from the $\partial_x^2$ term. The corresponding boundary dispersion is
\begin{align}
\omega = 2ck - 2ic\Gamma^2 k^2 -\frac{c\ell_B^2}{2}k^3 +\mathcal{O}(k^4),
\end{align}
which highlights the dispersive and diffusive nature of the chiral boson mode.

{\bf Charge Diffusion and KPZ:} Carrying the analysis of the previous section beyond linear response, we arrive at the following extension of Eq. (\ref{eq:diffusion}):
\be\label{edgedis} \p_t \rho + 2c\p_x \rho - D\p_x^2\rho+\gamma\p_x(\rho^2)+\p_x\eta=0\ee
where $D=2c\Gamma^2$ is the diffusion constant, and $\gamma$ denotes the leading nonlinearity. Note that $\rho$ is understood to be the Chiral boson edge charge evaluated at the boundary $\rho_{\text{CB}}(x,0,t)$, and we are only including the leading terms in the gradient expansion. The last term is a thermal noise contribution satisfying $\langle \eta(t,x)\eta(0,0)\rangle=2T\chi D\delta(t)\delta(x)$, which is required by virtue of the fluctuation-dissipation theorem, where $T$ is the temperature and $\chi$ is the edge charge susceptibility. We recognize (\ref{edgedis}) as the stochastic Burgers equation \cite{delacretaz2020breakdown}, whose correlations at long time and length scales are known to belong to the Kardar-Parisi-Zhang (KPZ) stochastic universality class \cite{kardar1986dynamic}.

A crucial characterizing feature of KPZ universality is that it is superdiffusive, despite what Eq. (\ref{edgedis}) suggests at the mean-field level, that we should find diffusive damping. This is a consequence of the presence of relevant interactions, i.e. the nonlinearity in (\ref{edgedis}) becomes strong at long wavelengths, as we now verify by performing power counting. Let us first assume that, at low frequencies and momenta, the scaling is governed by the mean-field linearized dynamics, which is diffusive: $\omega\sim k^z$, with $z=2$. Additionally, the noise correlation function tells us that $\eta \sim (\omega k)^{1/2}$. Balancing the linear terms we then find that the density scales as $\rho\sim k^{1/2}$. The interaction term thus is more relevant than the linear terms at long distances, i.e. $\p_x^2\rho\ll \p_x(\rho^2)$, implying that the dynamical scaling behavior cannot be diffusive. 

It is in fact possible to extract the correct dynamical scaling exponent $z$. First, one can verify that the stochastic process (\ref{edgedis}) admits the probability distribution $\mathcal P[\rho]\propto \exp\left(-\frac 1{2\chi T}\int dx\, \rho ^2\right)$ as an exact stationary distribution  \cite{kamenev2023field}, thus implying that $\rho\sim k^{1/2}$ exactly. Additionally, Eq. (\ref{edgedis}) possesses an emergent Galilean boost symmetry: $x\to x+2\gamma\epsilon t$, $\rho\to \rho+\epsilon$, for arbitrary $\epsilon$. This ties together the dimension of the first term and the nonlinear term, which immediately implies that $z=3/2$, i.e. the scaling dynamics is \emph{superdiffusive}.

{\bf Discussion:} In this work, we use the hydrodynamic framework to introduce dissipative effects near the edge of Laughlin states. When energy dissipation is present, the effective charge dynamics exhibit diffusive behavior in the linear regime. However, this behavior breaks down at long distances, where the interaction terms become more relevant.  By introducing nonlinearities and fluctuations, we observe the emergence of superdiffusive behavior, which is a fundamental characteristic of KPZ universality and potentially holds valuable insights regarding the edge physics of Laughlin states.
Notably, an earlier experiment conducted on GaAs \cite{talyanskii1994experimental} has already reported observations of dissipative scaling on the Quantum Hall edge, describing it as ``somewhere between linear and quadratic." It would be intriguing to revisit this observation using current experimental techniques.

Although the presence of two chiral modes propagating in the same direction may be related to a chiral central charge of $c_-=2$, we showed that the Kelvin mode does not undergo diffusion in the presence of dissipation. This observation strongly suggests that the Kelvin mode does not contribute to heat transport. This heuristic justification arises when considering edge dissipation as an alternative approach to studying heat transport. Therefore, the chiral central charge of the Laughlin state remains $c_-=1$, indicating that the chiral boson mode aligns with the physics of the FQH edge. Exploring the fate of the coastal Kelvin mode through the lens of condensed matter transport experiments remains an intriguing direction for future research.

{\bf Acknowledgments: } SG would like to thank Sarang Gopalakrishnan for useful discussions. SG and DR are supported by NSF CAREER Grant No. DMR-1944967. GMM was supported in part by the National Science Foundation under Grant OMA1936351. PG is supported by the Alfred P. Sloan Foundation through Grant FG-2020-13615, the Department of Energy through Award DE-SC0019380, and the Simons Foundation through Award No. 620869. Part of this work was performed at the Aspen Center for Physics, which is supported by National Science Foundation grant PHY-1607611.

\bibliographystyle{refstyle}
\bibliography{edgedissipation-Bibliography.bib}

\begin{thebibliography}{10}

\bibitem{girvin1987off}
S.~Girvin and A.~H. MacDonald.
\newblock \emph{Off-diagonal long-range order, oblique confinement, and the
  fractional quantum Hall effect}.
\newblock Physical review letters, \textbf{58}, 1252 (1987).

\bibitem{helium-book}
D.~Vollhardt and P.~W{\"o}lfle.
\newblock \emph{The superfluid phases of helium 3}.
\newblock Courier Corporation (2013).

\bibitem{Abrikosov1957TheMP}
A.~A. Abrikosov.
\newblock \emph{The magnetic properties of superconducting alloys}.
\newblock Journal of Physics and Chemistry of Solids, \textbf{2}, 199--208
  (1957).

\bibitem{stone1990superfluid}
M.~Stone.
\newblock \emph{Superfluid dynamics of the fractional quantum Hall state}.
\newblock Physical Review B, \textbf{42}, 212 (1990).

\bibitem{zhang1989effectivetheory}
S.~C. Zhang, T.~H. Hansson, and S.~Kivelson.
\newblock \emph{Effective-Field-Theory Model for the Fractional Quantum Hall
  Effect}.
\newblock Phys. Rev. Lett., \textbf{62}, 82--85 (1989).

\bibitem{zhang1992csgl}
S.~C. Zhang.
\newblock \emph{The Chern--Simons--Landau--Ginzburg theory of the fractional
  quantum Hall effect}.
\newblock International Journal of Modern Physics B, \textbf{6}, 25--58 (1992).

\bibitem{read1989orderparameter}
N.~Read.
\newblock \emph{Order Parameter and Ginzburg-Landau Theory for the Fractional
  Quantum Hall Effect}.
\newblock Phys. Rev. Lett., \textbf{62}, 86--89 (1989).

\bibitem{abanov2013fqhe}
A.~G. Abanov.
\newblock \emph{On the effective hydrodynamics of the fractional quantum Hall
  effect}.
\newblock Journal of Physics A: Mathematical and Theoretical, \textbf{46},
  292001 (2013).

\bibitem{monteiro2022topological}
G.~M. Monteiro, V.~Nair, and S.~Ganeshan.
\newblock \emph{Topological fluids and FQH edge dynamics}.
\newblock arXiv preprint arXiv:2203.06516 (2022).

\bibitem{monteiro2023coastal}
G.~M. Monteiro and S.~Ganeshan.
\newblock \emph{Coastal Kelvin Mode and The Fractional Quantum Hall Edge}.
\newblock arXiv preprint arXiv:2303.05669 (2023).

\bibitem{geracie2014effective}
M.~Geracie and D.~T. Son.
\newblock \emph{Effective field theory for fluids: Hall viscosity from a
  Wess-Zumino-Witten term}.
\newblock arXiv preprint arXiv:1402.1146 (2014).

\bibitem{monteiro2021hamiltonian}
G.~M. Monteiro, A.~G. Abanov, and S.~Ganeshan.
\newblock \emph{Hamiltonian structure of 2D fluid dynamics with broken parity}
  (2021).

\bibitem{Note1}
In the pure Chern-Simons (CS) framework of the FQH edge the bulk has no
  dynamics, and the CS term in the bulk only fixes the edge algebra through
  gauge invariance. The edge dynamics are subsequently added by hand~\cite
  {wen1990compressibility}. In contrast, within the first-order hydro
  framework, the bulk and boundary dynamics are intricately tied together by
  the no-stress boundary conditions, manifesting boundary layer structure and
  naturally resulting in a dynamical chiral boson theory that is coupled to the
  matter density at the edge.

\bibitem{kane1994randomness}
C.~Kane, M.~P. Fisher, and J.~Polchinski.
\newblock \emph{Randomness at the edge: Theory of quantum Hall transport at
  filling $\nu$= 2/3}.
\newblock Physical review letters, \textbf{72}, 4129 (1994).

\bibitem{kane1995impurity}
C.~Kane and M.~P. Fisher.
\newblock \emph{Impurity scattering and transport of fractional quantum Hall
  edge states}.
\newblock Physical Review B, \textbf{51}, 13449 (1995).

\bibitem{delacretaz2020breakdown}
L.~V. Delacr{\'e}taz and P.~Glorioso.
\newblock \emph{Breakdown of diffusion on chiral edges}.
\newblock Physical Review Letters, \textbf{124}, 236802 (2020).

\bibitem{kardar1986dynamic}
M.~Kardar, G.~Parisi, and Y.-C. Zhang.
\newblock \emph{Dynamic scaling of growing interfaces}.
\newblock Physical Review Letters, \textbf{56}, 889 (1986).

\bibitem{forster1977large}
D.~Forster, D.~R. Nelson, and M.~J. Stephen.
\newblock \emph{Large-distance and long-time properties of a randomly stirred
  fluid}.
\newblock Physical Review A, \textbf{16}, 732 (1977).

\bibitem{vznidarivc2011spin}
M.~{\u{Z}}nidari{\u{c}}.
\newblock \emph{Spin transport in a one-dimensional anisotropic Heisenberg
  model}.
\newblock Physical Review Letters, \textbf{106}, 220601 (2011).

\bibitem{wei2022quantum}
D.~Wei, A.~Rubio-Abadal, B.~Ye, F.~Machado, J.~Kemp, K.~Srakaew, S.~Hollerith,
  J.~Rui, S.~Gopalakrishnan, N.~Y. Yao, et~al.
\newblock \emph{Quantum gas microscopy of Kardar-Parisi-Zhang superdiffusion}.
\newblock Science, \textbf{376}, 716--720 (2022).

\bibitem{das2020nonlinear}
A.~Das, K.~Damle, A.~Dhar, D.~A. Huse, M.~Kulkarni, C.~B. Mendl, and H.~Spohn.
\newblock \emph{Nonlinear fluctuating hydrodynamics for the classical XXZ spin
  chain}.
\newblock Journal of Statistical Physics, \textbf{180}, 238--262 (2020).

\bibitem{Note2}
Since the stress tensor $T_{ij}$ is symmetric, we have used $T_{xy}=T_{yx}$.

\bibitem{Note3}
In fluid dynamics, the no-penetration condition is typically unchanged
  regardless of the dissipative nature of the fluid. Thus, we do not change the
  no-penetration condition in order to introduce dissipation.

\bibitem{levitov2016electron}
L.~Levitov and G.~Falkovich.
\newblock \emph{Electron viscosity, current vortices and negative nonlocal
  resistance in graphene}.
\newblock Nature Physics, \textbf{12}, 672--676 (2016).

\bibitem{kamenev2023field}
A.~Kamenev.
\newblock \emph{Field theory of non-equilibrium systems}.
\newblock Cambridge University Press (2023).

\bibitem{talyanskii1994experimental}
V.~Talyanskii, M.~Simmons, J.~Frost, M.~Pepper, D.~Ritchie, A.~Churchill, and
  G.~Jones.
\newblock \emph{Experimental investigation of the damping of low-frequency edge
  magnetoplasmons in GaAs-Al x Ga 1- x As heterostructures}.
\newblock Physical Review B, \textbf{50}, 1582 (1994).

\bibitem{wen1990compressibility}
X.~Wen and A.~Zee.
\newblock \emph{Compressibility and superfluidity in the fractional-statistics
  liquid}.
\newblock Physical Review B, \textbf{41}, 240 (1990).

\end{thebibliography}


\onecolumngrid

\section{Appendix: Fluid Equations from CSGL Action \label{app:csgl}}
\label{app:csgl}
In this section, we show how the first-order hydro equations for the FQH start are obtained by varying the CSGL action.
\begin{align}
S_{\text{CSGL}}&=\int d^3x\left[i\hbar\Phi^\dagger D_t\Phi - \frac{\hbar^2}{2m}\left|D_i\Phi\right|^2-V(|\Phi|^2) - \frac{\hbar\nu}{4\pi}\epsilon^{\mu\lambda\kappa} a_\mu\p_\lambda a_\kappa\right] ,
\end{align}
where $D_\mu=\partial_\mu+i\left(\frac{e}{\hbar}A_\mu+a_\mu\right)$,
gives rise to the fluid equations (\ref{eq:chargecon})-(\ref{eq:tij}). Varying this action with respect to the fields $(\Phi, \Phi^\dagger, a_\mu)$ gives
\begin{align}
&\delta S_{\text{CSGL}} = \int d^3x \Bigg\{ \left[ i\hbar  D_t\Phi + \frac{\hbar^2}{2m}D_i^2\Phi - V'(|\Phi|^2)\Phi\right]\delta\Phi^\dagger +\left[ -i\hbar  D_t\Phi^\dagger + \frac{\hbar^2}{2m}(D_i^2\Phi)^\dagger - V'(|\Phi|^2)\Phi^\dagger\right]\delta\Phi \nonumber \\
& + \left[ -\hbar|\Phi|^2 - \frac{\hbar \nu}{2\pi}\epsilon_{ij}\partial_i a _j\right]\delta a_0  + \left[ \frac{i\hbar^2}{2m}\left(\Phi^\dagger D_k\Phi-(D_k\Phi)^\dagger\Phi \right) -\frac{\hbar \nu}{2\pi}\epsilon_{ki}\left( \partial_i a _0 -\partial_t a _i \right)\right]\delta a_k\Bigg\} .
\end{align}
Note here that we have discarded all boundary terms arising from integrating by parts. From this, we get the following equations of motion
\begin{align}
& i\hbar  D_t\Phi + \frac{\hbar^2}{2m} D_i^2\Phi - V'(|\Phi|^2)\Phi =0, \label{eq:var1}\\
& |\Phi|^2 + \frac{ \nu}{2\pi}\epsilon_{ij}\partial_i a _j =0 , \label{eq:var2}\\
& \frac{i\hbar}{2m}\left[\Phi^\dagger D_k\Phi-(D_k\Phi)^\dagger\Phi \right] -\frac{ \nu}{2\pi}\epsilon_{ki}\left( \partial_i a _0 -\partial_t a _i \right) =0 , \label{eq:var3}
\end{align}
where the variation over $\Phi$ produced a redundant equation. To pass to a fluid description we write $\Phi = \sqrt{n}e^{i\theta}$, and define the fluid velocity as $v_i=\frac{\hbar}{m}\left(\p_i\theta+a_i+\frac{e}{\hbar} A_i+\frac{\epsilon_{ij}}{2n}\p^jn\right)$. With this we immediately see that Eq. (\ref{eq:var2}) enforces the Hall constraint 
\begin{align}
n-\frac{\nu e B}{2\pi\hbar}+\frac{\nu}{4\pi}\partial_i\left(\frac{\partial_i n}{n}\right) + \frac{\nu m}{2\pi\hbar}\epsilon_{ij}\partial_i v_j =0 ,
\end{align}
where we've defined the magnetic field as $B=\epsilon_{ij}\partial_i A_j$. On the other hand, Eq. (\ref{eq:var1}) can be split into a real and imaginary part yielding two independent equations. First we write the derivative terms in more detail,
\begin{align}
D_t\Phi &= \left[\frac{1}{2n}\partial_t n + i \partial_t\theta + i\left(\frac{e}{\hbar}A_0 + a_0\right) \right]\Phi ,\\
D_i^2\Phi &= \Bigg[\frac{m}{\hbar}\epsilon_{ij}\partial_i v_j - \epsilon_{ij}\partial_i\left(\frac{e}{\hbar}A_j + a_j\right) + i\frac{m}{\hbar}\partial_i v_i - \frac{m^2}{\hbar^2}v_i v_i +\frac{m}{\hbar n}\epsilon_{ij}v_i\partial_j n +i\frac{m}{\hbar n}v_i\partial_i n \Bigg]\Phi ,
\end{align}
where we've replaced $\theta$ in favor of the hydro variables whenever possible using the definition of $v_i$. The imaginary part of Eq. (\ref{eq:var1}) gives
\begin{align}
 0 &= \frac{\hbar}{2 n}\partial_t n + \frac{\hbar}{2}\partial_i v_i + \frac{\hbar}{2 n} v_i \partial_i n  , \\
 0 &= \partial_t n + \partial_i \left(n v_i\right) ,
\end{align}
which is precisely the continuity equation. To recover momentum conservation (\ref{eq:momcon}) takes more work. To start, we apply $-\frac{1}{\hbar}\partial_i$ to the real part of Eq. (\ref{eq:var1})
\begin{align}
\partial_t\partial_i\theta + \partial_i\left(\frac{e}{\hbar}A_0 + a_0\right) + \frac{1}{2}\epsilon_{jk}\partial_i\partial_k v_j + \frac{\hbar}{2m}\epsilon_{jk}\partial_i\partial_k \left(\frac{e}{\hbar}A_j + a_j\right) + \frac{m}{\hbar}v_j\partial_i v_j - \partial_i\left(\frac{1}{2n}\epsilon_{jk} v_j \partial_k n \right) + \frac{1}{\hbar} \partial_i V'(n) = 0 . \label{eq:longp}
\end{align}
Now, from the definition of $v_i$ we have the following two relations
\begin{align}
\partial_t\partial_i\theta &= \frac{m}{\hbar}\partial_t v_i - \partial_t\left(\frac{e}{\hbar}A_i + a_i\right) + \epsilon_{ij}\partial_j\left(\frac{1}{2n} \partial_k \left(n v_k \right)\right) , \\
v_j \partial_i v_j &= v_j \partial_j v_i +\frac{\hbar}{m}\epsilon_{ij}\epsilon_{kl}v_j\partial_k\left(\frac{e}{\hbar}A_l + a_l +\frac{1}{2n}\epsilon_{lp} \partial_p n\right) 
\end{align}
which upon substituting into (\ref{eq:longp}) gives a lengthy expression, but after some manipulations can be put in the form
\begin{align}
&\partial_t v_i + v_j\partial_j v_i + \frac{1}{m n}\partial_j \left[\frac{\hbar n}{2}\left(\epsilon_{ik}\partial_k v_j + \epsilon_{jk}\partial_i v_k\right)\right] + \frac{1}{m} \partial_i V'(n) +\frac{\hbar}{m}\partial_i \left(\frac{e}{\hbar}A_0 + a_0\right) - \frac{\hbar}{m}\partial_t\left(\frac{e}{\hbar}A_i+ a_i\right)  \nonumber\\
& + \frac{\hbar^2}{2m}\epsilon_{jk}\partial_i\partial_k\left(\frac{e}{\hbar}A_j + a_j\right) + \frac{\hbar}{m}\epsilon_{ij}\epsilon_{kl}v_j\partial_k\left(\frac{e}{\hbar}A_l + a_l\right) =0\,.
\end{align}
From here we note that Eq. (\ref{eq:var3}) can be written in terms of the hydro variables as
\begin{align}
n v_i - \frac{\hbar}{2m}\epsilon_{ij}\partial_i n &= \frac{\nu}{2\pi}\left(\epsilon_{ij}\partial_t a_j - \epsilon_{ij}\partial_j a_0  \right) , \\
\frac{2\pi}{\nu}\left( \epsilon_{ij}n v_j +\frac{\hbar}{2m} \partial_i n \right) &= \partial_i a_0 - \partial_t a_i,
\end{align}
which along with the definition of $B$ and the Hall constraint simplifies the last 4 terms
\begin{align}
\partial_t v_i + v_j\partial_j v_i &+ \frac{1}{m n}\partial_j \left[\frac{\hbar n}{2}\left(\epsilon_{ik}\partial_k v_j + \epsilon_{jk}\partial_i v_k\right)\right] + \frac{1}{m} \partial_i V'(n) + \frac{eB}{m}\epsilon_{ij}v_j + \frac{2\pi\hbar^2}{\nu m} \partial_i n ,
\end{align}
where we have assumed now that $\partial_i A_0 - \partial_t A_i=0$. Finally, note that
\begin{align}
\partial_i V'(n) = -\frac{1}{n}\partial_i\left[ V(n)-nV'(n) \right] ,
\end{align}
and so we recover Eq. (\ref{eq:momcon}) with stress tensor $T_{ij}$ given by (\ref{eq:tij}).


\end{document}